\documentclass[a4paper,12pt,onecolumn]{article} 

\usepackage[dvips]{graphicx}

\newcommand{\figuresize}{0.8\textwidth}
\newcommand{\BBox}{\sqcup\!\!\!\!\sqcap }

\begin{document} 
 
\begin{flushleft}
\Large
{\bf Fluctuation spectrum and size scaling of river flow and level\\}
\renewcommand{\thefootnote}{\fnsymbol{footnote}}
\large

{\bf Kajsa Dahlstedt and Henrik Jeldtoft Jensen\footnote[1]
{Author for correspondence (h.jensen@imperial.ac.uk; 
URL: http://www.ma.imperial.ac.uk/$\sim$hjjens)} \\}

\normalsize
{\it Department of Mathematics, Imperial College London, South Kensigton
campus, London SW7 2AZ, U.K. \\}
\renewcommand{\thefootnote}{\arabic{footnote}} 
 
\end{flushleft}
\linespread{1}

\begin{abstract} 
\noindent
We describe the statistical properties of two large river systems: the Danube and the 
Mississippi. The properties of the two rivers are compared qualitatively to the 
general properties of a critical steady state system. Specifically, we test
the recent suggestion by Bramwell, Fennell, Holdsworth and Portelli [{\it Europhys. Lett.}
{\bf 57}, 310 (2002)] that a universal probability density function (PDF) 
exists for the fluctuations in river level, namely the Bramwell-Holdsworth-Pinton
(BHP) PDF. 
The statistical properties investigated in this paper are: the PDF of the river 
flow and river level; moment scaling with basin area; moment to moment scaling or 
relative scaling;  and power spectral properties of the data. We find that the moments 
of the flow scale approximately with basin area and that 
the seasonally adjusted flows exhibit relative moment scaling. 

Compared to the Mississippi, the Danube shows large deviations from spatial scaling and 
the power spectra show considerable dependence on system size. This might be due to water 
use and regulations as well as inhomogeneities in the basin area.  
 We also find that the PDF of level data in some, but not all, cases can be 
 qualitatively approximated by the BHP PDF. We discuss why this coincidence
 appears to be accidental.
 
\end{abstract}


\section{Introduction}
The spatial structure of river basins and the statistics and dynamics of
the flow of rivers have been analysed by a number of authors using methods
from statistical mechanics. Rinaldo found that river basins are spatial
fractals \cite{Rinaldo:97}. The
flow of rivers were found to exhibit scaling \cite{Thomas:1970} and 
multiscaling \cite{Gupta:1990} with basin area for basins ranging from $10$ to $10^3$ ${\rm km}^2$.
Temporal correlations, when analysed by means of the power spectrum, 
are also considered to be multiscaling or multifractal with respect to
time \cite{Gupta:1990,Pandey:1998,Tessier:1996}. Perhaps the most remarkable
link to systems in statistical mechanics is an observation by Bramwell, Fennell
and Portelli \cite{Bramwell:2002}. They found  that the probability density function(PDF)
of the fluctuations of river level at 
Nagymaros, which is on the Danube, is in good 
agreement with a PDF found for the magnetic fluctuations in the classical XY-model close to 
criticality \cite{Bramwell:2000dec}. We will refer to this PDF as the Bramwell-Holdsworth-Pinton 
(BHP) PDF. The data from the Nagymaros was published in \cite{Janosi:1999} and the 
fit proposed by Bramwell et al. was checked visually in \cite{Bramwell:2002}.

The BHP PDF has been described and characterised by Bramwell et. al. \\
\cite{Bramwell:2000dec}. 
It has also been found in one experimental system and in computer simulations of SOC-models 
\cite{Bramwell:1998,Bramwell:2000april},\\ \cite{Dahlstedt:2001,Pinton:1999,Sinha-Ray:2001}.  
These investigations indicated that the BHP PDF describes fluctuations in many very 
different systems. It was suggested that the BHP PDF might be  universal and ubiquitous
in nature arising in systems which lack a characteristic length scale and exhibit finite 
size scaling similar to  critical equilibrium systems. 

Inspired by these observed links to statistical mechanics and scale invariance,
we want to address the following questions: Do the statistics of river
flow and level height typically exhibit scaling and, if so, what type of scaling? Is the observation of the BHP fluctuation spectrum at Nagymaros representative
of fluctuations of river lever in general and, if so, can the observed 
BHP fluctuation spectrum be related to scale free aspects of the river system?
It is not clear {\it a priori} which quantity one should look for when
searching for the BHP fluctuation spectrum 
\cite{Bramwell:2000dec,Bramwell:1998,Bramwell:2000april,Dahlstedt:2001,Pinton:1999,Sinha-Ray:2001}.
We will therefore analyse level as well as flow data.

We approach these issues by analysing data collected from several
stations, corresponding to different basin areas, along the rivers Danube 
and Mississippi, see table \ref{Table:Rivers}. We also consider data from one very small river, the Wye at Cefn Brwyn with a basin area of $10.6$ ${\rm km}^2$ \cite{NWA} and one very large river, the Rio Negro at Manaus with a basin area of $3\times10^6$  ${\rm km}^2$ \cite{Richey:1989}. 
This allows us to cary out a finite size scaling
analysis similar to that used in statistical mechanics to study
the scale free behaviour in systems with a diverging correlation length
\cite{Cardy:1990}.

It is useful to explain what we mean by finite size scaling. In equilibrium critical 
systems  one expects that the standard deviation $\sigma$ and 
the mean of the order parameter $\langle m\rangle$ scale with system size (close to criticality) such that a 
rescaled PDF becomes 
independent of system size. This is expressed by the hyperscaling relation
$\langle m\rangle \propto \sigma \propto L^\alpha$  
and the finite size scaling (FSS) hypothesis  \cite{Cardy:1990,Aji:2001}.

The FSS hypothesis can generally be written as 
\begin{equation}
P(m)=L^{-\beta} f(mL^{-\alpha},\frac{L}{\xi})
\label{eq:ScalingHypothesis}
\end{equation} 
 which holds if the system is in a critical state \cite{Bramwell:2000dec,Aji:2001}. 
If the spatial correlation length $\xi$ is infinitely larger than L, 
i.e. $\frac{L}{\xi}\to 0$, then we can make the approximation 
\begin{equation}
P(m)=L^{-\beta} f(mL^{-\alpha}). 
\end{equation}

The FSS hypothesis leads to a scaling of the moments which we can write as
\begin{equation}
\langle m^{n} \rangle \propto L^{n\theta}
\label{eq:ScalingMoments}
\end{equation}
where $\theta=\alpha$. This will be referred to as simple scaling.
The simple scaling of the moments can be extended to multiscaling by letting the 
moments of the order parameter $m$ scale as
\begin{equation}
\langle m^{n} \rangle \propto L^{n\theta(n)}.
\label{eq:MultiScaling}
\end{equation}
This extension of moment scaling is incompatible with the simple FSS hypothesis and 
with hyperscaling.

The  FSS hypothesis does not hold in the following three cases: the correlation 
length does not diverge and corrections due to the finite length of $\xi$ in relation
 to the system size have to be accounted for; the moments are multiscaling and the 
 hypothesis needs to be changed; more than one scaling field exists in the system, such that 
\begin{equation}
\langle m \rangle \propto L_1^{\alpha_1}  L_2^{\alpha_2} \ldots L_N^{\alpha_N}.
\label{eq:Scalingfileds}
\end{equation}
In the last case,  the FSS hypothesis can hold approximately if there is one dominant 
scaling field and all the other scaling fields are irrelevant \cite{Forgacs:91}.

\begin{table}[htb]
\begin{center}
\begin{tabular}{llll}
\large\bf{Mississippi}& &\large\bf{Danube}&\\
\bf{Name}& \bf{Basin area}&\bf{Name}&\bf{Basin area}\\
Royalton &30 040 ${\rm km}^2$ \cite{USGS}&Dillingen &11 315 ${\rm km}^2$ \cite{BAFG}\\
St. Paul &95 300 ${\rm km}^2$ \cite{USGS}&Ingolstadt&20 001 ${\rm km}^2$\cite{BAFG}\\
Clinton  &221 700 ${\rm km}^2$ \cite{USGS}&Nagymaros &183 534 ${\rm km}^2$ \cite{VITUKI}\\
Keokuk &308 200 ${\rm km}^2$ \cite{USGS}&Orsova &576 232 ${\rm km}^2$ \cite{GRDC}\\
Memphis &2 416 000 ${\rm km}^2$ \cite{USGS}&Ceatal Izmail &807 000 ${\rm km}^2$ \cite{GRDC}\\
\end{tabular}
\end{center}
\label{Table:Rivers}
\caption{River sites used }
\end{table}

The paper is structured in the following way. We begin by considering river
flow. In Secs. 2, 3 and 4 we discuss the PDF, the scaling and multiscaling
analysis and the power spectrum analysis of river flow. In Sec. 5 we present
the PDF of the fluctuations in the river level and Sec. 6 contains a discussion
and our conclusions.  

To avoid misunderstandings we mention that we use the following notation.
The river flow is denoted by $q$ and the height of the river level by $h$.
The seasonally adjusted flow and height (defined in the appendix) is denoted
by $s_q$ and $s_h$ respectively. 

\section{River Flow}
In this section  we present the comparison of the PDFs of seasonally adjusted 
flow data with varying basin area from the sites listed in table 
\ref{Table:Rivers}. The signal we consider is equal to the deviation 
from the average, over all years, at a specific day, say 5th of febuary, 
normalised by the standard deviation of that particular day. 
We defer details to the appendix.
This seasonally adjusted signal has by
construction a zero mean and a standard deviation equal to one.
Thus no further scaling is needed in order to plot the PDF of
the seasonally adjusted data in the form used previously to check
for Gaussian and BHP like behaviour, see e.g. \cite{Bramwell:1998,Bramwell:2000april}.

\begin{figure}[!htb]
\begin{center}
\includegraphics[width=\figuresize]{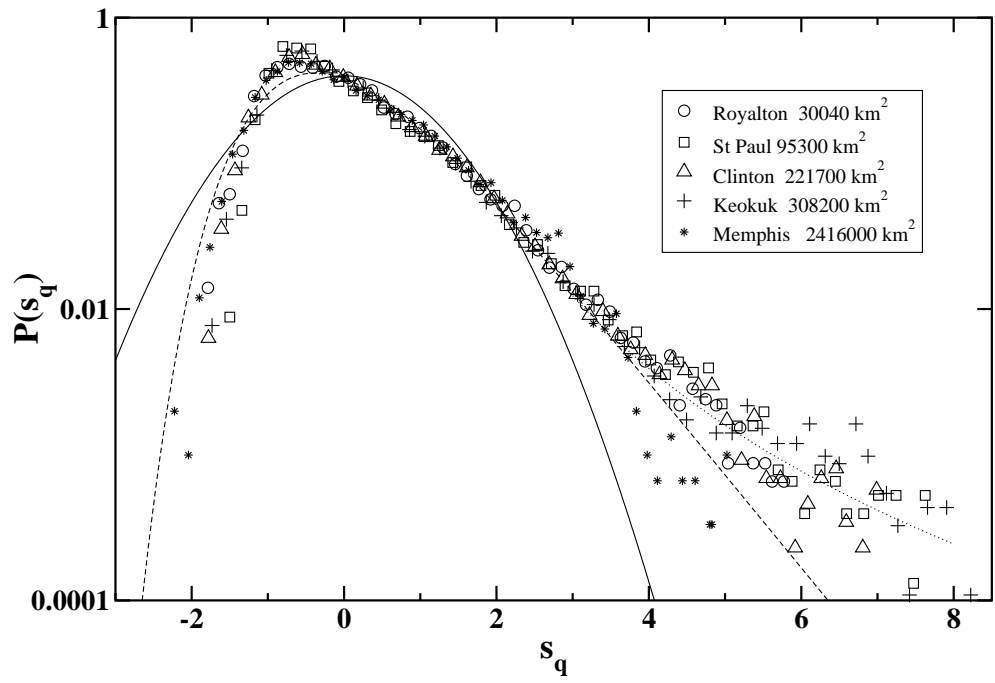}
\caption{The PDF of the flow for the five sites along the river Mississippi on a semi-log plot.
For reference, a normal PDF (solid line), a BHP PDF (dashed line) and a power-law  tail 
(dotted line) have been plotted. }
\label{mississippi}
\end{center} 
\end{figure}

Fig.~\ref{mississippi} shows the 
PDF of the seasonally adjusted flow for the river Mississippi. 
We plot the seasonally adjusted flow along the x-axis and the logarithm
of the PDF along the y-axis.
A normal distribution will appear as a parabola in this plot.
We note that the tails of the distributions vary 
slightly but there is an approximate data collapse. 
Hence adjusting for seasonal variation cancels the dependence on system size.

\begin{figure}[!htb]
\begin{center}
\includegraphics[width=\figuresize]{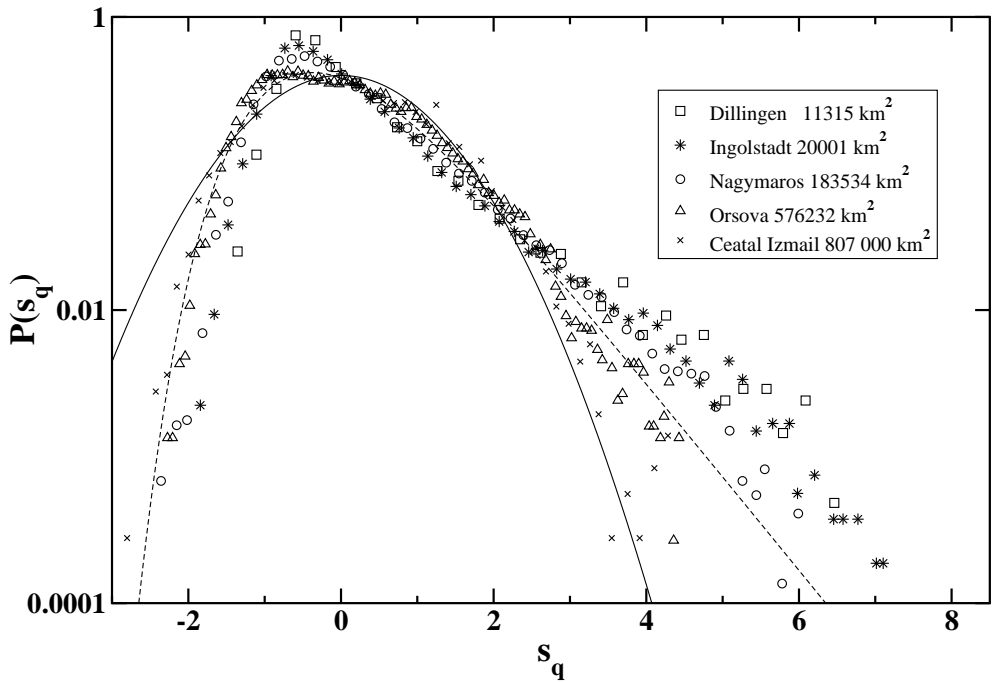}
\caption{The PDF of the flow for the five sites along the Danube on a semi-log plot. 
For reference, a 
normal (solid line) and a BHP (dashed line) PDF have been plotted. }
\label{danube}
\end{center} 
\end{figure}

In Fig.~\ref{danube} we show a similar plot for the Danube. In this plot, 
we can see a definite change of the PDF shape from the small basin area site, 
at Dillingen, to the large basin area site at  Ceatal Izmail. The shift in PDF is towards a 
Gaussian and it suggests that the spatial correlation length of the system is shorter than 
the system size at the largest site leading to a sum of uncorrelated contributions
to the flow. 
From the PDF analysis we expect the hyperscaling relationship and simple scaling to hold for 
the river Mississippi but not for the Danube. We also note that the
PDF of the fluctuations in the flow deviates strongly  
from the BHP PDF, indicated by the dashed line in both figures.
As we shall see in Sec. 5,  the situation is more subtle for river level
data.
    
\begin{figure}[!htb]
\begin{center}
\includegraphics[width=\figuresize]{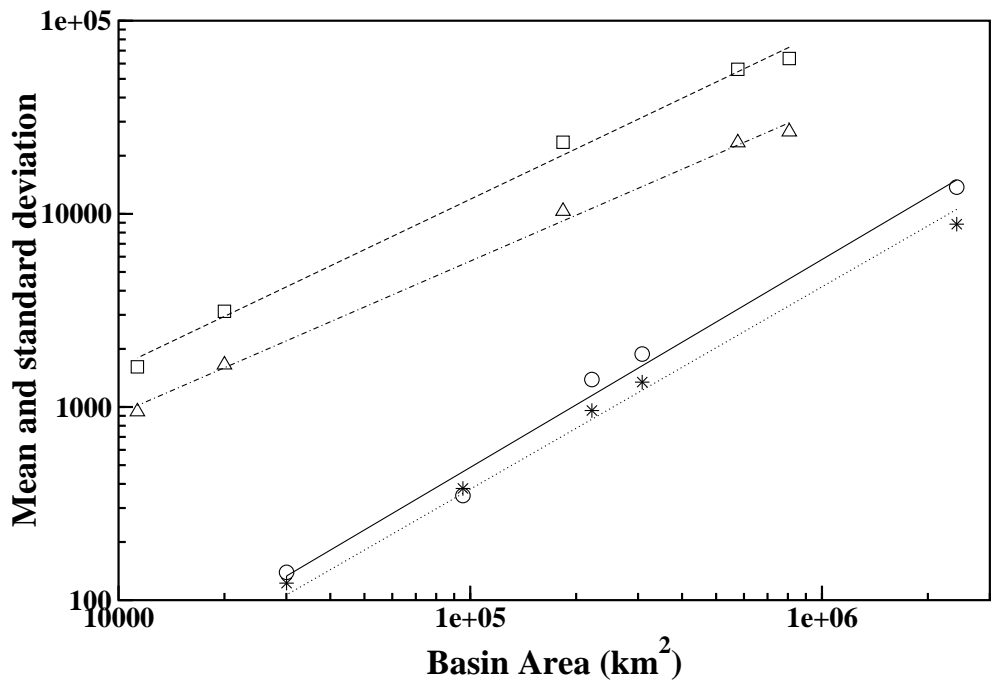}
\caption{The scaling of mean and standard deviation $\sigma$ with basin area on a 
log-log plot. River Mississippi mean [$\protect\circ$] and  $\protect\sigma$ 
[$*$]. River Danube mean [$\protect\BBox$] and $\protect\sigma$ 
[$\protect\triangle$]. The dashed line is the fit {$\langle q \rangle \propto 
L^{0.86779}$},the dashed-dotted line is the fit $\sigma_q \propto L^{0.78994}$, 
the solid line is the fit  $\langle q\rangle \propto L^{1.0763}$   and the dotted 
line is the fit $\sigma_q \propto L^{1.0477}$. The data from the Danube has been 
shifted with a factor of 10 for clarity.}
\label{Scaling:AreaMeanStd}
\end{center}
 
\end{figure}

Let us now consider the mean and the standard deviation of the ``raw''
data for which no adjustment for seasonal variation has been carried out.    
Fig.~\ref{Scaling:AreaMeanStd} compares the scaling of the mean with the basin 
area, $L$,  and the scaling of the standard deviation with $L$. 
The mean and 
standard deviation scale approximately as 
$\langle q\rangle \propto L^{1.08}$ and $\sigma_{q}\propto L^{1.05}$ for the Mississippi and for 
the river Danube approximately as $\langle q\rangle \propto L^{0.87}$ and 
$\sigma_{q}\propto L^{0.79}$. The exponents for $\langle q\rangle$ and
$\sigma_q$ are close, i.e. the hyperscaling 
relationship is approximately fulfilled for both rivers, though with greatest
accuracy for the Mississipi data.  

The analysis shows that the data collapse of the Mississippi
flow data 
is in fairly good agreement with the FSS  hypothesis of equation 
\ref{eq:ScalingHypothesis} and simple scaling of the first two moments.
The Danube has a larger deviation from the hyperscaling relationship, which we also 
expected from the analysis of the PDFs. This deviation might be due to a spatial 
decorrelation of the process, as we mentioned earlier. There may be several reasons 
for the decorrelation, such as inhomogeneities in the basin area and human regulation 
of the river.

\section{Scaling of higher moments and multiscaling}

\begin{figure}[!htb]
\begin{center}
\includegraphics[width=\figuresize]{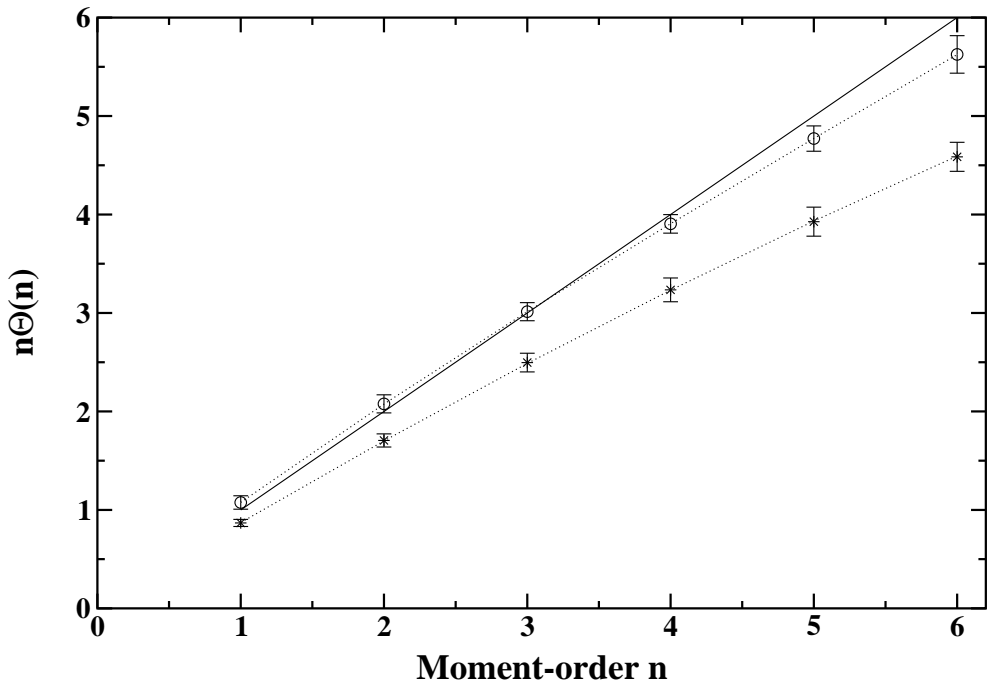}
\caption{Multiscaling function $n\Theta(n)$. The solid line is the simple scaling behaviour: 
the Danube [$*$],$\Theta (n) = 0.89278 - 0.021216 n$ and the river Mississippi  
[$\circ$], $\Theta(n) = 1.1198 - 0.046448 n + 0.0026748n^2$. The error bars shown are 
the errors from the regression analysis.}
\label{Thetanfunction}
\end{center} 
\end{figure}

We now  present data on the scaling of the higher moments of
the flow for the 
Mississippi and the Danube. We have shown that both rivers approximately fulfil 
the hyperscaling relationship. However, both rivers show deviations from simple scaling 
that look like multiscaling. The scaling exponents can be expressed as a function of moment-order $n$. 
This function is shown in Fig.~\ref{Thetanfunction}. 

\begin{figure}[!htb]
\begin{center}
\includegraphics[width=\figuresize]{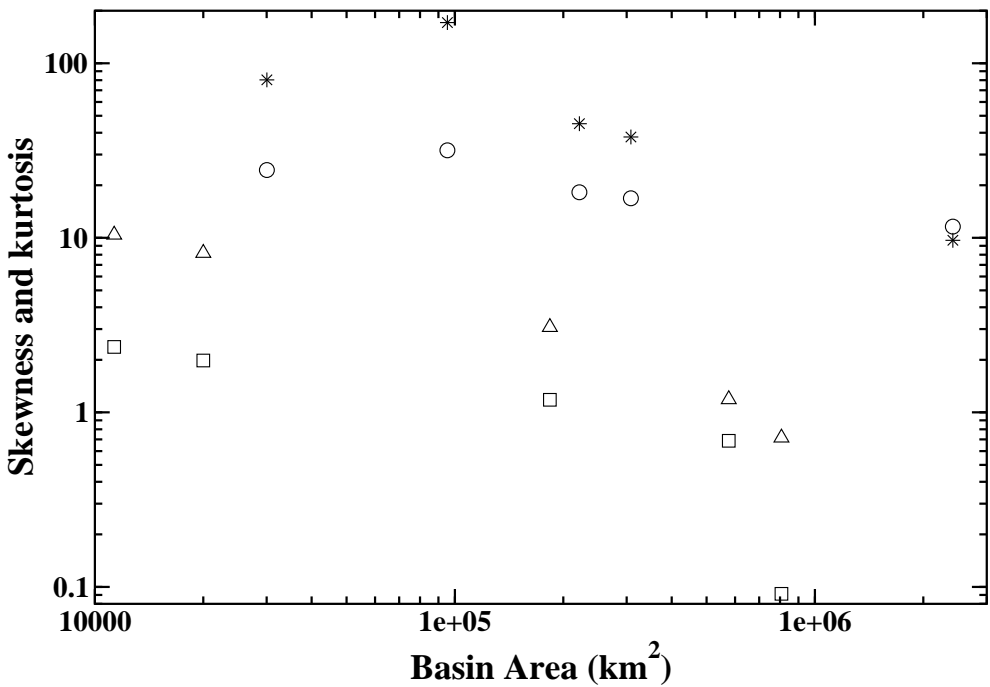}
\caption{Scaling of skewness and kurtosis to basin area on a log-log plot.  
River Mississippi skewness[$\protect\circ$] and  kurtosis[$*$]. River Danube 
skewness[$\protect\BBox$] and kurtosis[$\protect\triangle$]. The data from the 
Danube has been shifted with a factor of 10 for clarity.}
\label{AreaSkewKurt}
\end{center} 
\end{figure}

We have also looked at the scaling of skewness $\langle (m- \langle m \rangle)^3 \rangle/\sigma^{3}$
and kurtosis $\langle (m-\langle m \rangle)^4 \rangle/\sigma^{4}-3$ of the data. 
 Given the analysis above and assuming  the data is multiscaling the skewness and kurtosis  
 should not be constants and they should scale with basin area. For both the river Mississippi and 
 the Danube, skewness and the kurtosis as functions of basin area do not fit a power-law, 
as  shown in Fig.~\ref{AreaSkewKurt}. Despite this, there is a simple power-law  
relationship between the skewness and kurtosis. This is given by $kurt\propto skew^{2.75}$ for 
the Mississippi and  $kurt\propto skew^{1.78}$ for the Danube. Thus both rivers have systematic 
corrections to simple scaling that certainly look like 
multiscaling but, since the higher rescaled moments do not scale with basin area, neither of 
the rivers multiscale according to the standard definitions. 

\begin{figure}[!htb]
\begin{center}
\includegraphics[width=\figuresize]{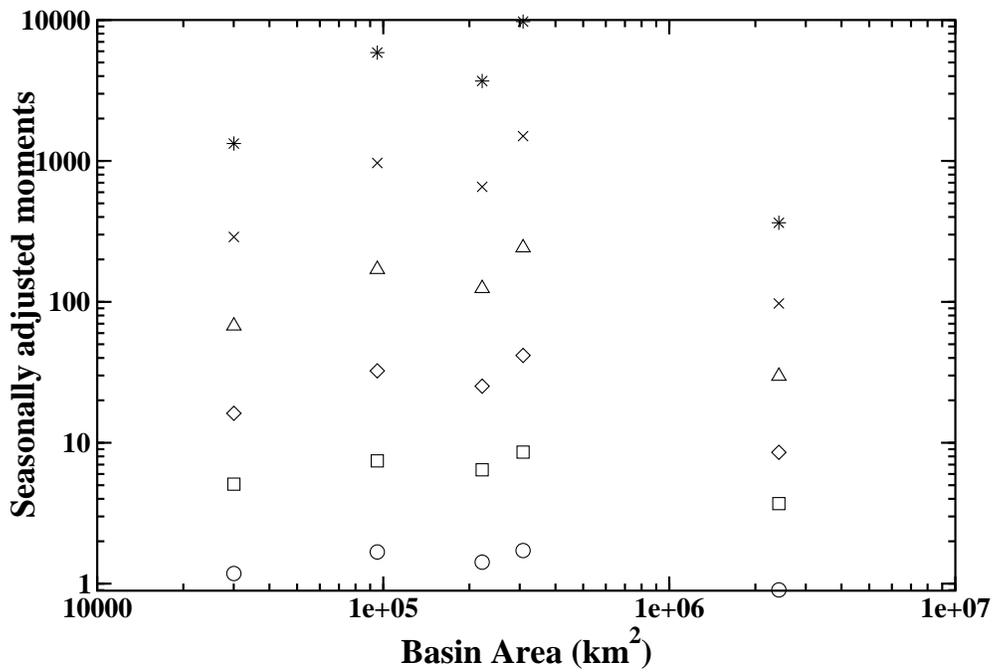}
\caption{Higher seasonally adjusted moments for the river Mississippi plotted 
against river basin area on a log-log plot. 3rd moment [$\circ$] to 8th moment [$*$]}
\label{arearelativescaling}
\end{center} 
\end{figure}

\begin{figure}[!htb]
\begin{center}
\includegraphics[width=\figuresize]{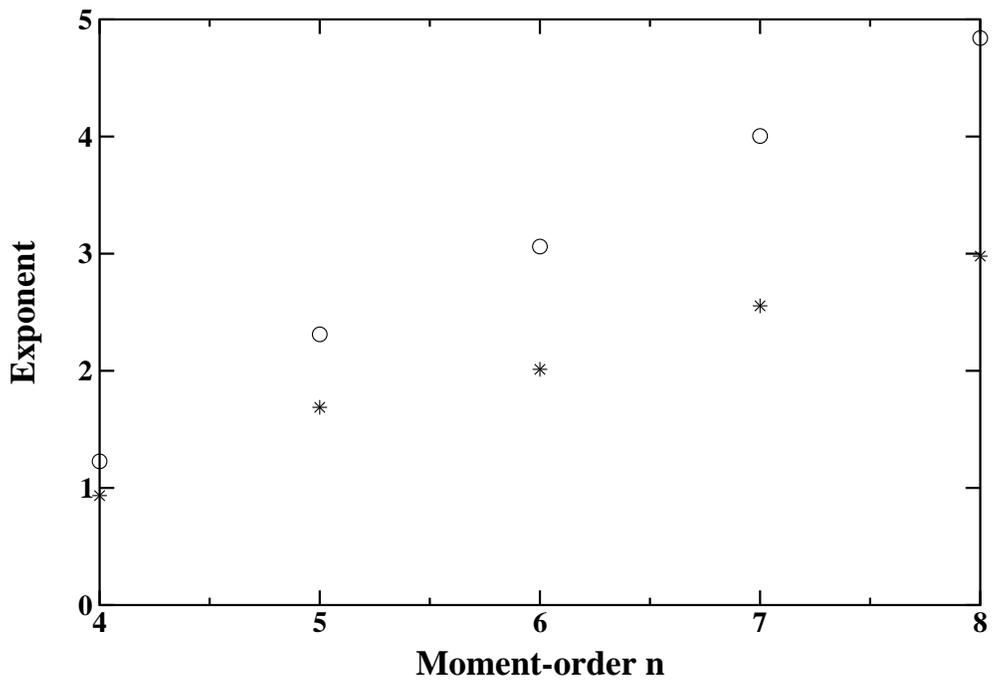}
\caption{The scaling of higher seasonally adjusted moments for the river  
Mississippi [$\circ$] and the river Danube [$*$].}
\label{relativescaling}
\end{center} 
\end{figure}

The deviations from multiscaling can also be seen in the seasonally adjusted data. The  
moments of the seasonally adjusted river Mississippi flow as a function of 
basin area do not fit a power-law as seen in Fig.~\ref{arearelativescaling}. 
Even so, the higher moments scale as power laws of the $3^{rd}$ moment with different exponents. 
We refer to this type of scaling as relative scaling of moments. 
The exponents are shown in Fig.~\ref{relativescaling} as functions of moment-order $n$.

We define the relative scaling as 
\begin{equation}
\langle m^{n} \rangle/\sigma^n \propto (\langle m^{n'} 
\rangle/\sigma^{n'})^{(n/n')K(n,n')}\mbox{ $n\neq n'> 2$}.
\label{eq:RelativeScaling}
\end{equation} 
Assuming that the moments follow simple scaling it is easy to see that $K(n,n')=0$,  and if the  
moments are multiscaling, relative scaling should be satisfied with the exponent $K(n,n')$ related to $\theta(n)$. 
So for  the Mississippi and the Danube we found that multiscaling
is not fulfilled and hence the observed relative scaling must have a different
origin.

Relative scaling is also possible if  we consider a weak secondary 
scaling field, such that the moments scale as 
\begin{equation}
\langle m^{n} \rangle = L_1^{n\alpha}L_2^{n\theta(n)}.
\label{eq:ScalingTwoFields}
\end{equation}
 If $\theta(2)\approx 0$ and  $\langle m  \rangle=0$, then $\sigma\propto L_1^{\alpha}$ 
 and $\langle m^{n}\rangle/\sigma^n\propto L_2^{n\theta(n)}$. In this way relative scaling 
 is possible even if $\langle 
 m^{n} \rangle/\sigma^n$ does not scale with $L_1$.
 For a related situation in equilibrium systems see \cite{Forgacs:91,FisherGelfand:88}.

The assumption of more than one scaling field is supported by a previous study of river basin 
characteristics. 
Regression analysis of different basin characteristics has shown that river flow is affected 
by many other features, in in addition to basin area, such as channel slope, precipitation and 
others \cite{Thomas:1970}. 

\section{Power Spectra of River Flow}
We now want to examine the signs of scale free behaviour from a temporal point
of view and study the long time correlations by means of  
 a power spectral analysis for the rivers 
Danube, Mississippi, Rio Negro and the Wye.  Generally the power spectra of rivers have 
two different frequency regimes  of power-law-like nature 
\cite{Tessier:1996}.  The two regimes are separated by a cross-over frequency $f_c$.  
Most river flow data is also considered to be multiscaling/multifractal with respect to 
time \cite{Gupta:1990,Pandey:1998,Tessier:1996}. In these studies data from different 
sites and rivers are considered to be realisations of the the same process and the power 
spectra are averaged over all sites and there are generally no references to how the 
cross-over frequency changes with basin area. Here we explicitly study
the dependence of the power spectrum on the basin area and we find that $f_c$ 
changes with basin area for the Danube but not for the Mississippi.
 
\begin{figure}[!htb]
\begin{center}
\includegraphics[width=\figuresize]{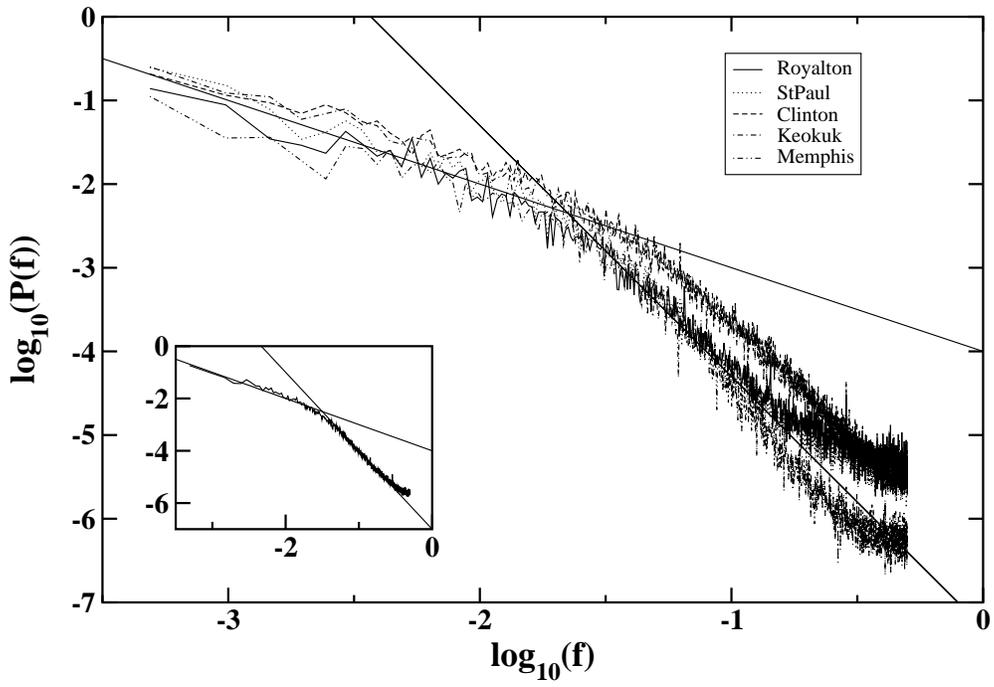}
\caption{Power spectra for the 5 sites along the Mississippi, for comparison  
two power-laws with exponents -1 and -3 are shown. 
The unit of frequency is inverse days.
The inset shows the averaged 
power spectra for all sites with approximate crossover of 30 days.}
\label{MississippiPower}
\end{center} 
\end{figure}

The Mississippi sites all have approximately the same spectrum which consists
of two power-law-like regimes separated by a cross-over frequency
 $f_c$ corresponding to approximately 20-40 days. In the low frequency domain the power-law  
has an exponent close to  $-1$ corresponding to logarithmic decaying correlations 
in the long time limit (see e.g. \cite{Jensen:98}) 
 and for high frequencies an exponent close to  $-3$, as can 
be  seen in Fig.~\ref{MississippiPower}. 
This indicates that the flow of the Mississippi
is determined by the intrinsic, system size independent, dynamics 
which produces time correlations that are longer than the longest observation time. 
Thus, within the  basin area size range investigated 
here we see critical-system-like behaviour: above we found that the PDF shows data collapse
and that the moments scale; in addition we observe that the power spectra are similar for all 
sites in the system and exhibit $1/f$ behaviour down to the lowest recorded
frequencies.

\begin{figure}[!htb]
\begin{center}
\includegraphics[width=\figuresize]{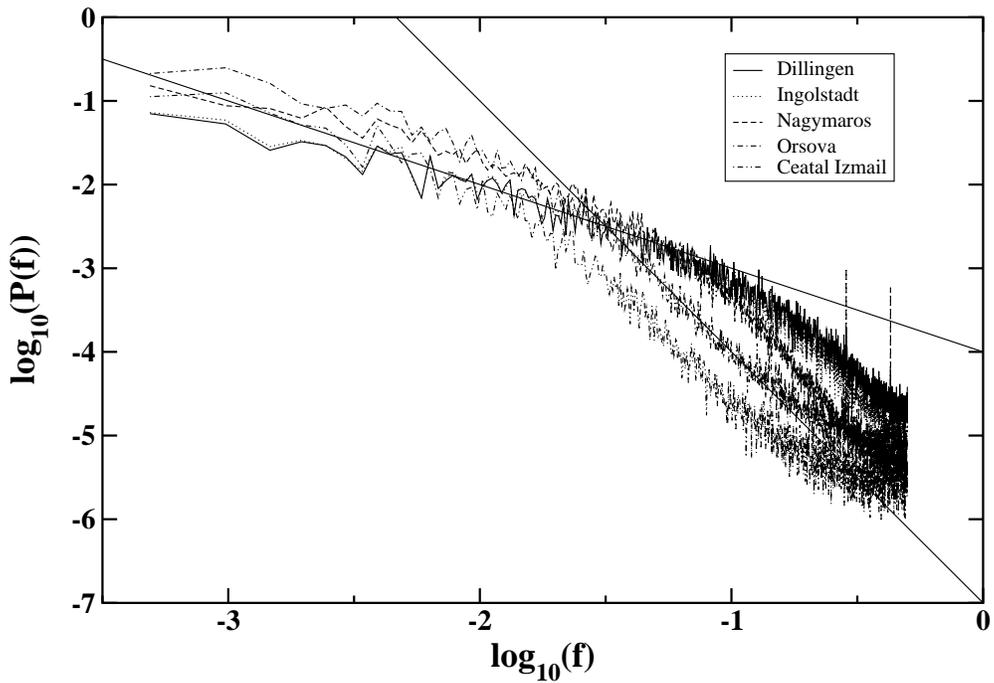}
\caption{Power spectra for the 5 sites along the Danube, for comparison 
two  power-laws with exponents -1 and -3 are shown
The unit of frequency is inverse days.}
\label{DanubePower}
\end{center} 
\end{figure}

For the Danube the crossover $f_c$ shifts with basin 
area with frequencies corresponding to approximately $1/f_c=60$ days for Cetal Izmail and 
Orsova, 20 days for Nagymaros and 10 days for Ingolstadt and Dilligen. 
In the low frequency region 
the spectrum does to a degree behave like $1/f$, see Fig.~\ref{DanubePower}. 
However, we suggest that the trend  
is that as the basin area is increased, the spectrum moves
towards a Lorentzian-like form, i.e $(1+(f/f_c)^2)^{-1}$. 
This would be indicative of exponentially decaying time correlations with a characteristic 
decay time $t_c=1/f_c$ and no essential time correlations beyond
$t_c$. The behaviour described here evince that the dynamics
of the Danube  is non critical and that its correlation time is
determined by the considered system size, or basin area.

\begin{figure}[!ht]
\begin{center}
\includegraphics[width=\figuresize]{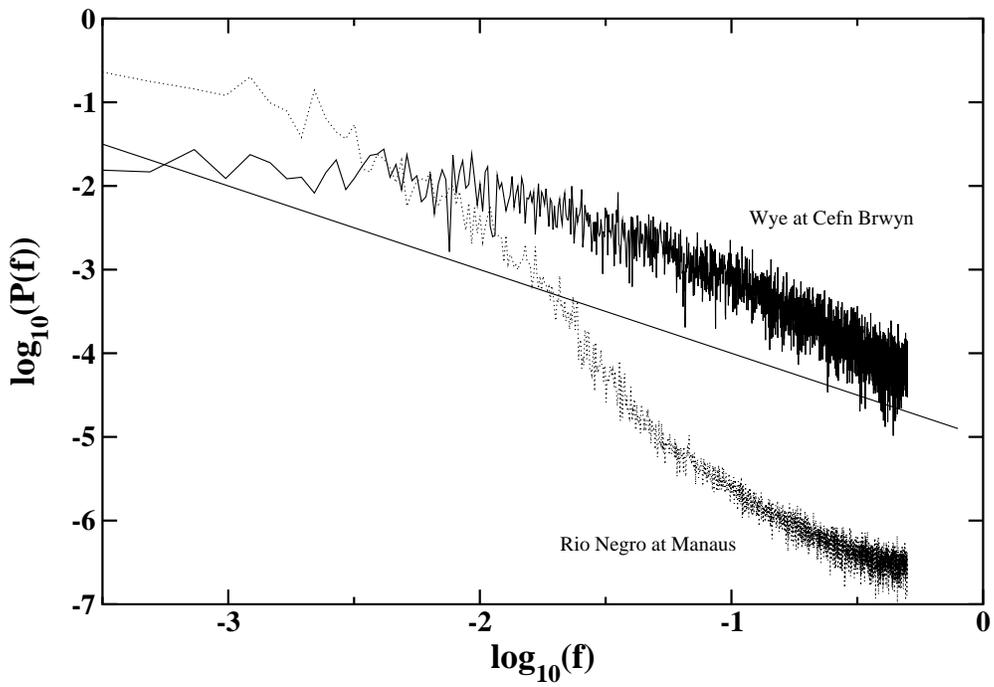}
\caption{Power spectra for the rivers Rio Negro at Manaus and Wye at Cefn Brwyn.
The straight line indicates $1/f$ behaviour. The unit of frequency is inverse days.}
\label{WyeAmazonPower}
\end{center} 
\end{figure}

We have also considered the flow of the Rio Negro at 
Manaus, with a very large basin area $3\times 10^6$ ${\rm km}^2$, and the flow of the Wye at 
Cefn Brwyn, with a very small basin area $10.6$ ${\rm km}^2$. The power spectra are  shown 
in Fig.~\ref{WyeAmazonPower}.

For the Rio Negro the spectrum seems to have a smooth transition between the low 
and the high frequency regime. The spectrum resembles, as in the case of 
the Danube at large basin areas, a Lorentzian form. Clearly it is not possible to identify accurately
the characteristic frequency, but a value of about $1/f_c=600$ days seems
reasonable. Thus the Rio Negro does not show critical power-law time
correlation, though its characteristic time is very long. If we assume that
long time corresponds to large spatial scales, our analysis
suggests that the Rio Negro is so large that its size exceeds the correlation
length of the river flow. The river therefore behaves as uncorrelated at the largest
spatial and temporal scales. 

The dynamics of the river Wye appears to be able to support $1/f$ correlations
down to a characteristic frequency $f_c$ corresponding to about 60 days. 
For frequencies lower than $f_c$ the spectrum becomes white,
thus the river system is so small that it cannot support correlations 
on time scales longer than these 60 days. 

\begin{figure}[!ht]
\begin{center}
\includegraphics[width=\figuresize]{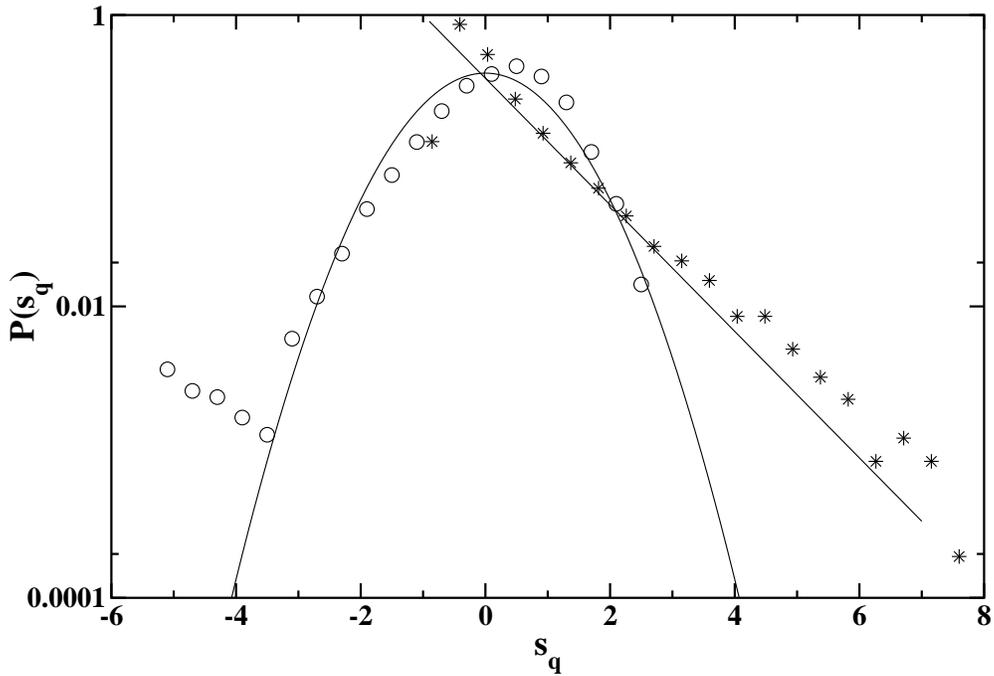}
\caption{Seasonally adjusted PDF for the rivers Rio Negro at Manaus[$\circ$] 
and Wye at Cefn Brwyn [$*$] on a semi-log plot, compared with an exponential 
and a Gaussian PDF. }
\label{WyeAmazonHistogam}
\end{center} 
\end{figure}

The PDFs of the flow of these two rivers are shown in Fig.~\ref{WyeAmazonHistogam}. There is a 
remarkable difference since the Wye has a nearly  exponential PDF and the Rio Negro has a 
Gaussian PDF. The Gaussian is consistent with our finding above that the
Rio Negro is so big that its dynamics are decorrelated. The exponential
tail found for the small river Wye is probably related to spatial correlations
extending up to the system size. This together with the $1/f$ behaviour
down to a certain frequency might reflect that the dynamics of the river Wye is critical, but since the system is rather small, the scale free range is
limited. 

\section{River-Level}
\begin{figure}[ht]
\begin{center}
\includegraphics[width=\figuresize]{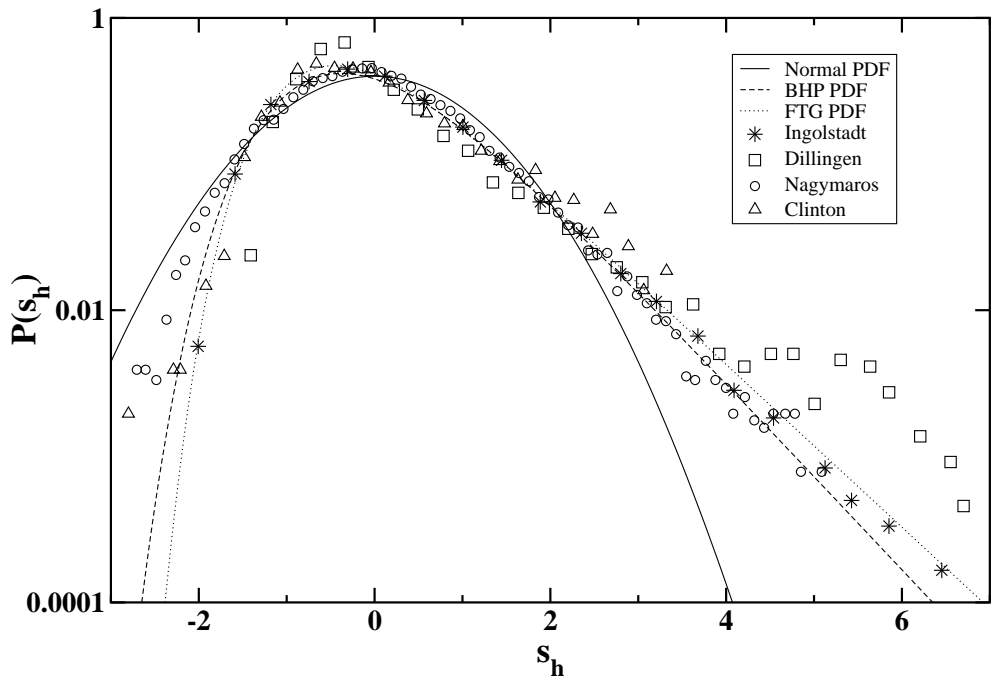}
\caption{The PDF of  river level for the sites Ingolstadt, Dillingen, Nagymaros 
and Clinton. As reference a normal (solid line), a BHP (dashed line) and a 
FTG (dotted line) PDF have been plotted. }
\label{levels}
\end{center} 
\end{figure}

In this section, we present the results of the analysis of river level. 
Fig.~\ref{levels} shows four PDFs of river level. The data is from the 
stations Clinton on the Mississippi, and  Dilligen, Ingolstadt and Nagyramos 
on the Danube. The seasonally adjusted PDFs are similar at the sites of Ingolstadt,
Nagyramos and Clinton and significantly different for the Dillingen site. The data 
that collapses seems to do so on or close to the BHP PDF but it is also close to a
 first order extreme value, Fisher-Tippet-Gumbel (FTG), PDF \cite{Gumbel:1958}. 
It is  difficult to determine visually from the histogram how good the fit is.

To get a better estimate of the fit, table  \ref{Level:highermoments} shows the 
seasonally adjusted higher moments of 
 the data compared to the moments of the BHP and a FTG PDF \cite{Bramwell:2000dec}. This shows that 
 quantitatively the fit is poor.
\begin{table}[!h]\begin{center}
\begin{tabular}{lllll}
Name/Moment& 3&4&5&6\\ 
FTG&1.14& 5.40&  18.57&91.39\\
BHP&0.8907& 4.415& -& -\\
Clinton&0.77& 3.05& 5.67&17.37\\
Nagymaros&0.54& 3.66&6.18& 27.31\\
Ingolstadt&0.76& 3.64&7.84& 28.44\\

\end{tabular}\end{center}
\label{Level:highermoments}
\caption{Seasonally adjusted moments $\langle s_h^n \rangle$ for three of the sites, 
the 1st order extreme value  PDF and the 
BHP PDF} 
 
\end{table}

\begin{figure}[!tb]
\begin{center}
\includegraphics[width=\figuresize]{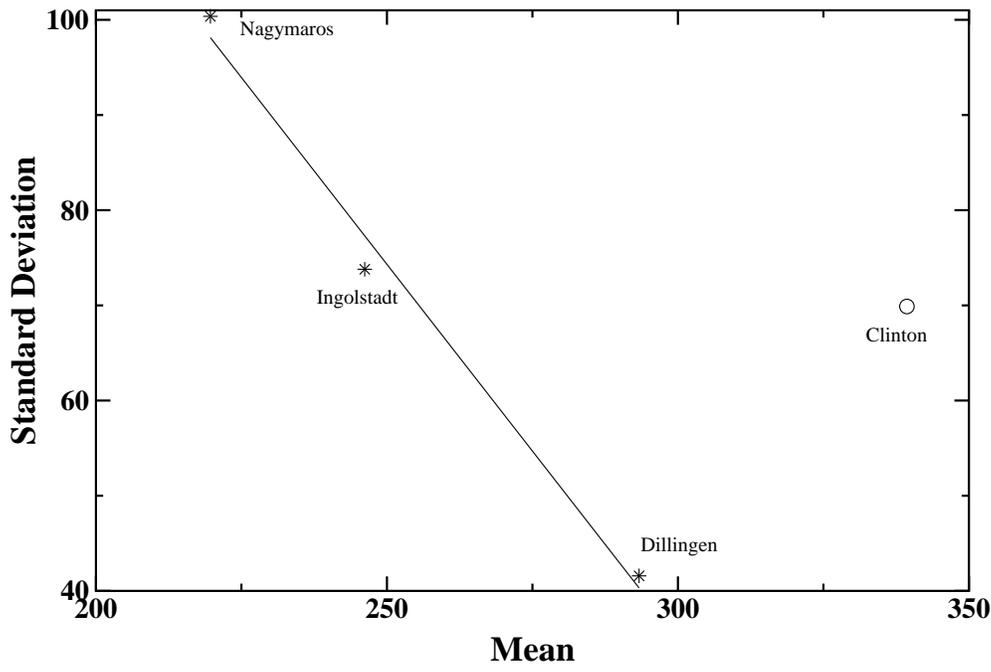}
\caption{The scaling of river level mean to standard deviation for the Danube [$*$] 
and Mississippi [$ \circ$].}
\label{Scaling:LevelMeanStd}
\end{center} 
\end{figure}

We also investigated whether the moments of a river level fulfil the hyperscaling
 relationship and scale with basin area. The relationship between the mean and the 
 standard deviation is shown in Fig.~\ref{Scaling:LevelMeanStd}. For the Danube we can make 
 an ad hoc linear fit, which does not suggest hyperscaling since 
 the slope is negative. From a scaling analysis with basin area (not shown here) 
 we  conclude that river level is not scaling with basin area. Thus the hyperscaling 
 and scaling is not the cause of the apparent data collapse of the data from the Ingolstadt 
 and Nagymaros sites. Although some of the data collapses close to the BHP, we can only 
 conclude that river level does not in general have a single universal PDF 
 upon rescaling. Furthermore, in the case where river level fluctuations are reasonably
 well described by the BHP form, the lack of scaling prevents us from concluding
 that the agreement with the BHP PDF arises as a consequence of underlying
 scale free behaviour. 


\section{Discussion and Conclusion} 
We have shown that PDFs of the fluctuations of the  river flow of the river Mississippi 
for different basin areas can all be approximately collapsed on to one functional
form by size scaling. However, this scaling function is not of the
BHP form.  In addition to the data collapse 
we have shown that the flow follows approximately simple scaling, with
deviations.  
In the case of the river Mississippi, the deviations resemble  multiscaling though
quantitative analysis reveals that the behaviour is inconsistent with 
multiscaling. We suggested a simple form with a weak second scaling field, independent of 
basin area which might explain the observed relative scaling of the seasonally adjusted moments.
In the case of the river Danube, the deviations from simple scaling have features in common with  
multiscaling. 
However, in this case the deviations were found to be related to spatial decorrelation between the 
sites along the river consistent with the lack of size scaling collapse of the
PDFs. Our conclusion is that the river Mississippi behaves like one correlated
whole whereas for the Danube, correlations are destroyed as the basin area is
increased.

We summarise that some degree of scale invariance can be found for rivers of 
intermediate size. Hence, our analysis confirms and extends previous observations of
scaling in smaller river systems \cite{Thomas:1970,Gupta:1990}.

We set out to determine whether  
measurements of river level fluctuations consistently have a universal PDF 
close to that of the BHP PDF.
For two out of the four river sites we investigated, the average daily level PDFs 
approximately collapse onto one PDF. This PDF resembles to the BHP PDF but quantitatively 
the fit is poor. The two other sites have a somewhat different shape of the PDF. 
We also found that there is no size scaling with basin area for river level. 
Hence even when the BHP form is found, it cannot be related to
an underlying scale free behaviour of the Danube. It would have
been very interesting to be able to analyse the level data for the Mississippi.
Unfortunately we could only obtain flow data for the Mississippi. Nevertheless,
our analysis suggests that fluctuations in river level do not in general
follow the BHP functional form. The observations in  \cite{Bramwell:2002}
appear to be accidental.

In addition to the analysis reported here we have investigated financial
data, weather data, tree ring data and other data and  could not find the BHP PDF. 
River level data appeared to be an exception rather than the rule.
Thus, in general we are compelled to conclude that the BHP PDF is rather difficult 
to find in nature. 

\section{Acknowledgement}
We thank Andrew Parry for enlightening and informative discussions.  
Kajsa Dahlstedt is supported by the foundation Helge Ax:son Johnson, the foundation 
BLANCEFLOR Boncompagni-Ludovisi, n{\'e}e Bildt and the EPSRC.
We are grateful to Paul Anderson and Gerald Law for linguistic
advise on the manuscript.

\section{Appendix}
The flow of a river is generally highly dependent upon the seasons. Therefore all the data has been 
seasonally adjusted with the seasonal mean and seasonal standard deviation 
denoted $\langle x\rangle_d$ and $\sigma_d$. Let $x(d,y)$ denote the measured
value of either the flow $q$ or the level $h$ measured at day number $d$ in year
number $y=1,\dots,N_y$. 
The adjusted data is defined as 
\begin{displaymath}
s_x=\frac{x(d,y)-\langle x\rangle_d}{\sigma_d}. 
\end{displaymath}
The seasonal mean and the seasonal standard deviation are defined 
as 
\begin{displaymath} 
\langle x\rangle_d=\frac{1}{N_y} \sum_{y=1}^{N_y}x(d,y)
\end{displaymath} and 
\begin{displaymath} 
\sigma_d=\sqrt{\frac{1}{N_y} \sum_{y=1}^{N_y}(x(d,y)-\langle x\rangle_d)^2}.
\end{displaymath}
The adjusted  data have $\sigma_s=1$ and $\langle s\rangle=0$ 
which can easily be checked.

\bibliography{river}

\begin{thebibliography}{}

\bibitem[NWA, 2002]{NWA}
 (2002).
\newblock National Water Archive, http://www.ceh.ac.uk/data/NWA.htm.

\bibitem[USG, 2002]{USGS}
 (2002).
\newblock US Geological Survey, http://www.usgs.gov/.

\bibitem[BAF, 2002]{BAFG}
 (2002).
\newblock Bundesanstalt f{\"u}r Gew{\"a}sserkunde Hydrologische Datenbank,
  http://www.bafg.de/.

\bibitem[VIT, 2002]{VITUKI}
 (2002).
\newblock VITUKI Rt. Institut of Hidrology, http://www.vituki.hu/.

\bibitem[GRD, 2002]{GRDC}
 (2002).
\newblock Global Runoff Data Centre, http://www.bafg.de/grdc.htm.

\bibitem[Aji and Goldenfeld, 2001]{Aji:2001}
Aji, V. and Goldenfeld, N. (2001).
\newblock Flucuations in finite critical and turbulent systems.
\newblock {\em Phys. Rev. Lett.}, 86(6):1007--1010.

\bibitem[Bramwell et~al., 2000a]{Bramwell:2000april}
Bramwell, S., Christensen, K., Fortin, J.-Y., Holdsworth, P., Jensen, H., Lise,
  S., Lopez, J., Nicodemi, M., Pinton, J.-F., and Sellitto, M. (2000a).
\newblock Universal fluctuations in correlated systems.
\newblock {\em Phys. Rev. Lett.}, 84(17):3744--3747.

\bibitem[Bramwell et~al., 2002]{Bramwell:2002}
Bramwell, S., Fennell, T., Holdsworth, P., and Portelli, B. (2002).
\newblock Universal fluctuations of the danube water level: A link with
  turbulence,criticality and company growth.
\newblock {\em Europhys. Lett.}, 57(3):310--314.

\bibitem[Bramwell et~al., 2000b]{Bramwell:2000dec}
Bramwell, S., Fortin, J.-Y., Holdsworth, P., Peysson, S., Pinton, J.-F.,
  Portelli, B., and Sellitto, M. (2000b).
\newblock Magnetic flucuations in the classical xy model: the origin of an
  exponential tail in a complex system.
\newblock {\em Phys. Rev. E}, 63(4):1106--1128.

\bibitem[Bramwell et~al., 1998]{Bramwell:1998}
Bramwell, S., Holdsworth, P., and Pinton, J.-F. (1998).
\newblock Universality of rare fluctuations in turbulence and critical
  phenomena.
\newblock {\em Nature}, 396:552--554.

\bibitem[Cardy, 1990]{Cardy:1990}
Cardy, J. (1990).
\newblock {\em Vol. 2 finite-Size Scaling, Current Physics Sources and
  Comments}.
\newblock North Holland, Amsterdam.

\bibitem[Dahlstedt and Jensen, 2001]{Dahlstedt:2001}
Dahlstedt, K. and Jensen, H.~J. (2001).
\newblock Universal fluctuations and extreme value statistics.
\newblock {\em J. Phys. A: Math. Gen.}, 34:11193--11200.

\bibitem[Fisher and Gelfand, 1988]{FisherGelfand:88}
Fisher, M.~E. and Gelfand, M.~P. (1988).
\newblock The reunions of three dissimilar vicious walkers.
\newblock {\em Journal of Statistical Physics}, 53:175--189.

\bibitem[Forgacs et~al., 1991]{Forgacs:91}
Forgacs, G., R.Lipowsky, and Nieuwenhuizen, T.~M. (1991).
\newblock The behaviour of interfaces in ordered and disordered systems.
\newblock In Domb, C. and Lebowitz, J., editors, {\em Phase Transition and
  Critical Phenomena}, volume~14. Academic Press, New York.

\bibitem[Gumbel, 1958]{Gumbel:1958}
Gumbel, E.~J. (1958).
\newblock {\em Statistics of Extremes}.
\newblock Columbia University Press, New York, NY.

\bibitem[Gupta and Waymire, 1990]{Gupta:1990}
Gupta, V.~K. and Waymire, E. (1990).
\newblock Multiscaling properties of spatial rainfall and river flow
  distributions.
\newblock {\em J. Geophys. Res. D.}, 95(3):1999--2009.

\bibitem[J{\'a}nosi and Gallas, 1999]{Janosi:1999}
J{\'a}nosi, I.~M. and Gallas, J.~A. (1999).
\newblock Growth of companies and water-level fluctuations of the river danube.
\newblock {\em Physica A}, 271:448--457.

\bibitem[Jensen, 1998]{Jensen:98}
Jensen, H.~J. (1998).
\newblock {\em Self-Organized Criticality}.
\newblock Cambridge University Press, New York, NY.

\bibitem[Pandey et~al., 1998]{Pandey:1998}
Pandey, G., Lovejoy, S., and D.Schertzer (1998).
\newblock Multifractal analysis of daily river flows including extremes for
  basins of five to two million square kilometres, one day to 75 years.
\newblock {\em J. Hydrology}, 208:62--81.

\bibitem[Pinton et~al., 1999]{Pinton:1999}
Pinton, J.-F., Holdsworth, P., and Labbe, R. (1999).
\newblock Power fluctuations in a closed turbulent shear flow.
\newblock {\em Phys. Rev. E}, 60(3):2452--2455.

\bibitem[Richey et~al., 1989]{Richey:1989}
Richey, J.~E., Nombre, C., and Deser, C. (1989).
\newblock Amazon river discharge and climate variability: 1903 to 1985.
\newblock {\em Science, New Series}, 246(4926):101--103.

\bibitem[Rodriguez-Iturbe and Rinaldo, 1997]{Rinaldo:97}
Rodriguez-Iturbe, I. and Rinaldo, A. (1997).
\newblock {\em Fractal river basins: chance and self-organization}.
\newblock Cambridge University Press, Cambridge, New York.

\bibitem[Sinha-Ray et~al., 2001]{Sinha-Ray:2001}
Sinha-Ray, P., de~{\'A}gua, L.~B., and Jensen, H. (2001).
\newblock Threshold dynanmics, multifractality and universal fluctuations in
  the soc forest-fire: facets of an auto-ignition model.
\newblock {\em Physica D}, 157:186--196.

\bibitem[Tessier et~al., 1996]{Tessier:1996}
Tessier, Y., shaun Lovejoy, Hubert, P., and Schertzer, D. (1996).
\newblock Multifractal analysis and modeling of rainfall and river flows and
  scaling, causal transfer functions.
\newblock {\em Journal of Geophysical Research}, 101:26427--26440.

\bibitem[Thomas and Benson, 1970]{Thomas:1970}
Thomas, D.~M. and Benson, M.~A. (1970).
\newblock Generalization of streamflow characteristics from drainage-basin
  characteristics.
\newblock {\em US Geological Survey, Water-Supply Paper}, (1975):75pp.

\end{thebibliography}
\newpage

\noindent{\Large \bf Figures}\\ 


\end{document}